\documentclass{svmult}
\usepackage{makeidx}     % allows index generation
\usepackage{graphicx}    % standard LaTeX graphics tool
\usepackage{multicol}    % used for the two-column index
\makeindex
\def\under{\,|\,}
\def\argmax{\mathop{\rm argmax}}

\begin{document}
\title*{Automated Resolution of Noisy Bibliographic References}
\author{Markus Demleitner\inst{1,2}, Michael Kurtz\inst{2}, 
Alberto Accomazzi\inst{2},
G\"unther Eichhorn\inst{2},
Carolyn S.~Grant\inst{2},  Steven
S.~Murray\inst{2}}

\institute{Lehrstuhl f\"ur Computerlinguistik der Universit\"at Heidelberg,
Karlstr.~2, 69117 Heidelberg, Germany
\and
NASA Astrophysics Data System, Harvard-Smithsonian Center for
Astrophysics, 60 Garden Street, Cambridge, MA 02138, USA}

\authorrunning{Demleitner, Kurtz, et al}
\maketitle
\begin{abstract}
We describe a system used by the NASA Astrophysics Data System to
identify bibliographic references obtained from scanned article pages by
OCR methods with records in a bibliographic database.  We analyze the
process generating the noisy references and conclude that the three-step
procedure of correcting the OCR results, parsing the corrected string
and matching it against the database provides unsatisfactory results.
Instead, we propose a method that allows a controlled merging of
correction, parsing and matching, inspired by dependency grammars.  We
also report on the effectiveness of various heuristics that we have
employed to improve recall.
\end{abstract}

\section{Introduction}

The importance of linking scholarly publications to each other
has received increasing
attention with the growing availability of such materials in
electronic form (see, e.g., van de Sompel, 1999).  The use of
citations is probably the most straightforward
approach to generate such links.

However, most publications and authors still do not give machine readable 
publication identifiers like
DOIs in their reference sections. The automatic generation of links
from references therefore is a challenge even for recent literature.
Bergmark (2000), Lawrence et~al.~(1999) and 
Claivaz et~al.~(2001) investigate methods to solve this problem under
a record linkage point of view.

For historical literature, the situation is even worse in that not
even the ``clean'' reference strings as intended by the authors 
are usually available.  In 1999,
the NASA Astrophysics Data System (ADS, see Kurtz et~al., 2000) began to
gather reference
sections from scans of astronomical literature and subsequently 
processed them with OCR software.
This has yielded about three million references (Demleitner et~al., 1999), 
many of them with severe recognition errors.  We will call 
these references
\emph{noisy}, whereas references that were wrong in the original
publication will be denoted \emph{dangling}.  Noisy references show
the entire spectrum of classic OCR errors in addition to the usual
variations in citation style.  Consider the following examples:

\begin{quote}
Bidelman, W. P. 1951, Ap. J. "3, 304; Contr. McDonald Obs., No.
199.\hfil\break
Eggen, 0. J. 195oa, Ap.J. III, 414; Contr. Lick Obs., Series II, No.
27.\hfil\break
---195ob, ibid. 112, 141; ibid., No. 30.\hfil\break
Huist, H. C. van de. 1950, Astrophys. J. 112,1.\hfil\break
8tro\char'176mgren, B. 1956, Astron. J. 61, 45.\hfil\break
Morando, B. 1963, "Recherches sur les orbites de resonance, "in Proceedings of t
he First International Symposium on the Use of Artilicial Satellites for Geodesy
, Washington, D. C. (North- Holland Publishing Company, Amsterdam, 1963), p. 42.
\end{quote}

While our situation was worse than the one solved by the record
linkage approaches cited above, we had the advantage of being able to
restate the problem into a classification problem, since we were only
interested in resolving references to publications contained in the ADS'
abstract database -- or decide that the
target of the reference is not in the ADS.  This is basically a
classification problem in which there are (currently) 3.5~million categories.

A method to solve this problem was recently developed by
Takasu (2003) using Hidden Markov Models to both parse
and match noisy references (Takasu calls them
``erroneous'').  While his approach is very different from ours,
we believe the ideas behind and our experiences with our resolver may
benefit other groups also facing the problem of resolution of noisy
references.

In the remainder of this paper, we will first state
the problem in a rather general setting, then discuss the basic ideas
of our approach in this framework, describe the heuristics we used to
improve resolution rates and their effectiveness and finally discuss
the performance of our system in the real world.

\section{Statement of the Problem}

\begin{figure}
\centering
\includegraphics{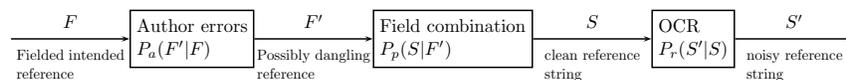}
\caption{A noisy channel model for references obtained from OCR. $F$
and $F'$ are a tuple-valued random variables, $S$ and $S'$ are
string-valued random variables.}
\label{noisychannel}
\end{figure}

Fig.~\ref{noisychannel} shows a noisy channel model for the generation
of a noisy reference from an original reference that 
corresponds to an entry in a bibliographic database.  In principle,
the resolving problem is obtaining 
\begin{equation}
\argmax_{F}P_m(F'\under
F)\,P_p(S\under F')\,P_r(S'\under S),\quad S\in\Sigma^\ast,
F'\in(\Sigma^\ast)^{n_f}
\label{argmax-eq}
\end{equation}for a given noisy reference $S'$.  Here,
$\Sigma$ is the base alphabet (in our case, we normalize everything to
7-bit ASCII) and $n_f$ is the number of fields in a bibliographic
record.  The domain of random variable $F$ is the database
plus the special value $\emptyset$ for references that are missing
from the data base but nevertheless valid.

The straightforward approach of modeling each distribution
mentioned above separately and trying to compute (\ref{argmax-eq})
from back to front will not work very well.  To see why, let us briefly
examine each element in the channel.  

Under a typical model for an OCR system, $P_r(S'\under S)$, will have many likely
$S$ for any $S'$, since references do not follow
common language models\footnote{To give an
example, the sequence ``L1'' will have a
very low probability in normal text, but, depending on the reference
syntax employed by authors, could occur in up to 2.5\% of the
references in our sample (it is actually found in 1.7\% of the OCRed
strings).} and are hard to model in general because of
mixed languages and (as text goes) high entropy.

In contrast, the ``parsing'' distribution $P_p(S\under F')$ 
is sharply peaked at few values.  Although reference syntax is much less
uniform than one might wish, even regular grammars can cope with a large
portion of the references, avoiding ambiguity altogether.  Even if
the situation is not so simple in the presence of titles or with
monographs and conferences, the number of interpretations for a given
value of $S$ with nonvanishing likelihood will be in the tens.

In the matching step modeled by $P_m(F'\under F)$, we have a similar
situation.  For journal references, ambiguity is very low indeed, and
even for books this record linkage problem is harmless
with $P_m(F'\under F)$ sharply peaked on
at worst a few dozen $F$.  The main complication here is detecting the
case $F=\emptyset$.

So, while $P_m$ and $P_p$ have quite low conditional entropies,
the one of $P_r$ is very high.  This is unfortunate, because in
computing (\ref{argmax-eq}) one would generate many $S$ only to throw
them away when computing $P_p$ or $P_m$.

In this light, an attempt to resolve noisy references along the lines
of Accomazzi et~al.~(1999)'s suggestion for clean references -- which
boils down to computing $\argmax_F P_m\left(F\under \argmax_{F'}
P_p(F'|S)\right)$ -- is bound to fail when extended to noisy
references.

It is clear that there have to be better ways since the
conditional entropy of $P(F\under S')$ is rather low, as can
be seen from the fact
that a human can usually tell very quickly what the correct
interpretation for even a very noisy reference is, at least when
equipped with a bibliographic search engine like the ADS itself.

Takasu (2003) describes how Dual and Variable-length
output Hidden Markov Models can be used to model a combined
conditional distribution $P_{p,r}(F'\under S')$, thus exploiting that
many likely values of $S$ will not parse well and therefore have a low
combined probability.  The idea of combining distributions is
instrumental to our approach as well.

\section{Our Approach}

\subsection{Core resolution}

One foundation of our resolver comes from 
dependency grammars (Heringer, 1993) in
natural language processing, which are based
on the observation that given the ``head'' of a (natural
language) phrase (say, a verb), certain
``slots'' need to be filled (e.g., eat will usually have to be
complemented with something that eats and something that is eaten).

In the domain of reference resolving, the equivalent of a phrase is
the reference.
As the head of this phrase, we chose the publication source, 
i.e., a journal or conference name, a book title, a
hint that a given publication is a Ph.D.~thesis or a preprint.  This
was done for three reasons.  Firstly, it is easy to
robustly extract this information from references in our domain,
secondly, there are relatively few possible heads (disregarding
monographs), and thirdly,
the publication source governs the grammar of the entire
reference.

For example, in addition to the publication year and
authors references to most journals 
need a volume and a page , while a
Ph.D.~thesis is complemented by a name of an institution, and
reports or documents from the ArXiv~preprint
servers may just take a single number.

Let us for now assume that references follow the regular
expression \emph{Author+ Year Rest}, where
Rest contains a mixture of alphabetic and numeric characters, and a
title is not given for parts of article collections
-- in astronomy, almost all references
follow this grammar.
A simple regular expression can identify the year with very close to 100\% recall
and precision even in noisy references, yielding a robust fielding of
the reference.

To find the head as defined above, we simply 
collect all alphabetic characters from the
Rest. The remaining numeric
information, i.e., all sequences of digits separated by non-digits,
are the fillers required by the head.  This exploits that 
fillers are almost always numeric and avoids dependency on syntactic
markers like commas that are very prone to misrecognition.  
Heads that have non-numeric fillers (mostly theses and monographs)
receive special treatment.

This head is matched against an authority file that
maps $N_t$ full titles and common
abbreviations for the sources known to the ADS to a ``bibstem''
(cf.~Grant et~al., 2000).  We select the $n$-best matching of these, where
$n=5$ proved a good choice.  
To assess the quality of a match, a string edit distance suffices.
The one we use is $$1-{(\Delta(a,h)-|a|)L(a,h)\over |h|},$$
where $a$ and $h$ are a string from the authority file and the head,
respectively, $\Delta(a,h)$ denotes the number of matching trigrams
from $h$ that are found in $a$, $L(a,h)$ is the plain Levenshtein
distance (Levenshtein, 1966) and $|\,.\,|$ is the length of the string.  
The worst-case runtime of this procedure is
$O(|h|\max(|a|)N_t\log N_t)$, but since we compute trigram
similarities first and compute Levenshtein distances only for
those $a$ having at least half as many trigrams in common with $h$ as
the best matching $a$,
typical run time will be of order
$O(|h|^2\log|h|)$.

This corresponds to maximizing
$P_{p,r}((\ldots,{\it source},\ldots)\under S')$, i.e., we derive a
distribution on publication sources directly from the noisy reference.
The conditional entropy of this distribution is relatively low, because 
there are few possible sources (order $10^4$) and the edit distance
induces a sharply peaked distribution.

For each bibstem, the number of slots and their
interpretation is known\footnote{Actually, we have an exception list
and normally assume two slots, volume and page.}, and we can
simply match the slots with the fillers or give educated guesses on
insertion or deletion errors based on our knowledge of the fillers
expected.  In the noisy channel model, this corresponds to greedily evaluating
$P_{p,r}(F'\under S',(\ldots,{\it
source},\ldots))$.  While in principle, the distribution would have a
rather high conditional entropy (e.g., many readings for the numerals
would have to be taken into account), it turns out that most of these
complications can be accounted for in the matching step, alleviating
the need to actually produce multiple $F'$, even more so since parsing
errors frequently resemble errors made by authors in assembling their
references, which are modeled in $P_a$.

If filling the slots with the available fillers is not possible,
the next best head is tried, otherwise, we have a complete fielded
record $f'$ that can be matched against the database using a $P_m$ to
be discussed shortly.  If this
matching is successful, the resolution process stops, otherwise, the
next best head is tried.

The matching has to be a fast operation since it is potentially tried
many times.  Fortunately, the bibliographic 
identifiers (bibcodes, see Grant et~al., 2000) used by the ADS are, for
serials, computable from the record in constant time, and thus,
matching requires a simple table lookup,
taking $O(\log N_r)$ time for $N_r$ records we have to match
against.

Due to the construction of bibcodes, the plain bibcode match only
checks the first character of the first authors' last name.  
The numbers below show that the entropy of
references with respect to the distribution implied by our algorithm
is so low that this shortcoming does not impact precision noticeably
-- put another way, the likelihood that OCR errors conspire to produce
a valid reference is very small even without using most of the
information from the author field.

The core resolving process typically runs in $O(\log N_r |h|^2\ln|h|)$ time.
On a 1400 MHz Athlon XP machine, a python script implementing this
resolves about 100~references per second and already catches more than
84\% of the total resolvable references in our set of 3,027,801
noisy references.

\subsection{Reference Matching}

For journals for which the database can be assumed complete $P_m(F\under F')$ 
is nontrivial, i.e., different from $\delta_{F,F'}$.  The single most
important ingredient is a mapping from volume numbers to publication
years and vice versa, because even if one field is wrong 
because of either OCR or author errors, the other can be
reconstructed.  We also scan the surrounding page range (authors
surprisingly frequently use the last page of an article) and try
swapping adjacent digits in the page number.  
Finally, we try special sections of journals
(usually letter pages).  The definition of this matching implies that
$P_m(F=f\under F'=f')=0$ if $f$ and $f'$ differ in more than one field.

While these rules are somewhat ad hoc, they are also
straightforward and probably would not profit from learning.
They alone account for 8\% of the successfully resolved references
without further source string manipulation.

When any of these rules are applied, the authors given in the
reference are matched against those in the data base using 
a tailored string edit distance.  It is computed by
deleting initials, first names and common
non-author phrases (currently ``and'', ``et'', and ``al'') and
then evaluating $${\it fault}=\sum_{w'\in A'}\min_{w\in A} 
L(w',w),$$ where $L$ is the Levenshtein distance with all weights one
and $A$ and $A'$ the author last names for the paper in the database and
from the reference, respectively.  The edit distance then is $d_a=1-{\it fault}/{\it
limit}$, where {\it
limit} is given by allowing 2 errors for each word shorter than 5
characters, 3 errors for each word shorter than 10 characters and 4
errors otherwise.  This reflects that OCR
language models do much better on longer words than on shorter ones,
even if they come from non-English languages.  Unless we have reason
to be stricter (usually with monographs), we accept a match if
$d_a>0$.

If, after all string manipulations described below have not yielded a
match, we relax $P_m$ for all sources
and also try to match identifiers with a different
first author (in case the author order is wrong), scan a page range of
plausible mis-spellings and try identifiers with different
qualifiers\footnote{This is necessary if there is more than one
article mapping to the same bibcode on one page, for details see
Grant et~al.~(2000).}.  7.8\% of the total
resolved references were only accepted after this.  We have not
attempted to ascertain how many of these references were dangling in
the original publication.

\subsection{Monographs and Theses}

The procedures described above are useful for serials and article
collections of all kinds.  Two kinds of publications have to be
treated differently.

As mentioned above, theses have alphabetic fillers.  
Thus, we use keyword spotting (a
hand-tailored regular expression for possible readings of ``Thesis'')
to identify the head within the rest.  Together with the first
character of the author's last name and the publication year, 
we select a set of candidates and
match authors and granting institutions analogous to the author
matching procedure described above.

Monographs are completely outside this kind of handling. For them, a
set of candidates is selected based on the first character of the
author name and the publication year, and authors and titles are
matched.  Since this is a very time-consuming procedure, it is only
attempted if the resolving to serials failed.

Note that using authors as heads as is basically done with
monographs would probably most closely mimic the techniques of 
human librarians. However, given
the fragility of author names both in the OCR
process and in transliteration, we doubt that a low-entropy
distribution would result from doing so.

\section{Heuristics}

Takasu (2003) conjectured that the comparatively unsatisfactory
performance of his method could be significantly improved through the
use of a set of heuristics.  We find that the same is true for our
approach. Almost 16\% of the total resolved papers only become
resolvable by the algorithm outlined above after some heuristic
manipulations are performed on the noisy reference.

We apply a sequence of such manipulations
ordered according to their ``daringness'' and re-resolve after each
manipulation.  These manipulations -- typically regular-expression
based string operations -- model a noisy channel, but of course
it would be very hard to write down its governing distribution.
Still, it may be useful to see what heuristics had what
payoff.

In a first step, we correct the most frequently
misrecognized abbreviations based on regular expressions for the
errors.  We concentrate on abbreviations because misrecognitions in
longer words usually do not confuse our matching algorithm.
While better models may have a higher payoff, our method
only contributes
0.6\% of the total resolved references.

The second step is more effective at 1.7\% of the total
resolved references.  We code rules about common misreadings of
numerals in a set of regular expressions, including substituting
numerals at the beginning of the reference using
a unigram model for OCR errors, fixing numerals within the reference string using a
hand-crafted bigram model and joining single digits to a preceding
group to make up for blank insertion errors.

At 4.9\% of the total still more effective are transformations 
on the alphabetic part behind the year, including
attempts to remove additional specifications (e.g., ``English
Translation''), and mostly very
domain-specific operations with the purpose of increasing the
conformity of journal specifications with the authority information
used by the source matcher.  The most important measure here, however,
is handling very short
publication names (``AJ'') that are particularly hard for the OCR.
From these experiences we believe a learning system will have to have a
special mode for short heads.

The last fixing step is dissecting the source specification along
separators (we use commas and colons) and try using the part that
yields the best match against the authority file
as the new head.  This usually removes bibliographic information
primarily in references to conference proceedings.  0.9\% of the total resolved
references become resolvable after this.  Note that this step would be
more important if we had to frequently deal with title removal.

Further, less interesting, heuristics are applied to bring references
into the format required by the resolver including title removal
-- for astronomy references,
this is rarely needed --, reconstruct references that refer to other reference's
parts, and to split reference lines containing two or more
references.  This last task
only applies to the rare entries consisting of two separate references
listed together by the author.  The resolver makes no attempt to discover
errors in line joining that were made earlier in the processing chain.

\section{Application}

Our dataset from OCR currently contains 3,027,801 references (some
$10^4$ of which actually consist of non-reference material
misclassified by the reference cutting engine).  Of these, 2,552,229
(or about 84\%) could be resolved to records in the database.

In order to assess recall and precision of the system described here,
we created a subset of 852 references
by selecting each reference with a probability of 0.00025, which yielded
118 references that were not resolved and 734 that were resolved. 
We then manually resolved each selected reference, correcting dangling
references as best we could.  Thus, the following numbers compare the
resolver's $P(F\under S')$ with a human's $P(F\under S')$.

The result was that two of the 734 resolved records were incorrectly
resolved.  In both cases, the correct record was not in the ADS, which
illustrates that the $F=\emptyset$ problem dominates the issue of
precision.
Of the non-resolved records, 94 were not in
our database, while 23 were, though six of these were marked doubtful
by the human resolvers.  Counting doubtful cases as errors, we
thus have a precision of more than 99\% and a recall of about 97\%.  
Of the 17 definite
false negatives, 7 are severely dangling or excessively noisy references to journals, while 6 are
references to conference proceedings and the rest monographs.

Note that it is highly unlikely that any of the drawn references were
ever inspected during the development of the heuristics.  Still, one
might question if evaluating the resolver with data that at least
might have been used to ``train'' it is justified.
Since during development we mainly inspected resolving 
failures rather than possibly incorrectly resolved references, we
would expect the fact the we did not hold back pristine reference data
for evaluation purposes to impact recall more than precision.

For journal literature between 1981 and 1998, we also compared 
the resolver result with data purchased
from ISI's science citation index\footnote{See http://www.isinet.com/}.
Randomly selecting 1\% of the articles covered by
ISI and removing references to sources outside the 
ISI sample, we had 10832 citing-cited
pairs, of which 311 were missing in the OCR sample and 1151 were
missing from ISI.

A manual examination of the citing-cited pairs missing from the OCR
sample revealed that 112
were really attributable to the resolver, 107 were due to incorrect
reconstructions of reference lines, and 86 references were missed because
the references were not found by the reference zone identification.

Of the references apparently missing from ISI, 2 were due to 
resolver errors\footnote{Actually, in one case the OCR conspired to
produce an almost valid reference to a wrong paper, in the second
case, incorrect line joining resulted in two references that were
mangled into a valid one.}, and less than 20\% were dangling references that
ISI did not correct, but were clearly identifiable nevertheless.  
We have not attempted to identify why the other (correct) pairs
were missing from our ISI sample; most problems probably
were introduced during the necessarily conservative matchup between
records from ISI and the ADS, and possibly in the selection of our data set
from ISI's data base.

For journal articles (others are, for the most part, 
not available from ISI), we can thus state a recall of 99\% and a
precision of 99.9\% for our resolver and a recall of about 97\% for
the complete system.

\section{Discussion}

In this paper we contend that robust interpretation of bibliographic 
references, as
required when resolving references obtained by current OCR techniques,
should integrate as much information obtainable from a set of known
publications as possible even in parsing and not delay incorporating
this information to a ``matching'' or linkage phase.

Our approach has been inspired by dependency grammars, in which a head
of a phrase governs the interpretation of the remaining elements.  For
(noisy) references, it is advantageous to use the name or type
of the publication as head.
The existence of
bibliographic identifiers that are for most references easily computable from
fielded records has been instrumental for the performance of our
system.

While we believe some of the rather ad hoc string manipulations and
edit distances employed by our current system can and should be
substituted by sound and learning algorithms, it seems evident to us
that a certain degree of domain-specific knowledge (most notably, a
mapping between publication dates and volumes) is very important for
robust resolving.

The system discussed here has been in continuous use at the ADS for
the past four years, for noisy references from OCR as well as for
references from digital sources.  The ADS
in turn is arguably the most important bibliographic tool in astronomy and
astrophysics.  The fact that the ADS has received very few complaints
concerning the accuracy of its citations backs the estimates 
of recall and precision given above.

\begin{acknowledgement}
We wish to thank Regina Weineck for help in the generation of validation
data.

The NASA Astrophysics Data System is funded
by NASA Grant NCC5-189.
\end{acknowledgement}

\end{document}